\documentclass[preprint, showpacs, reprintnumbers, amsmath, amssymb, superscriptaddress]{revtex4-2}
\usepackage{graphicx}
\usepackage{amsmath}
\usepackage{amssymb}
\usepackage{epstopdf}
\usepackage{amsmath}

\usepackage{gensymb}
\usepackage{color}
\usepackage{hyperref}
\usepackage{soul}
\hypersetup{
    colorlinks=true,
    citecolor=red,
    linkcolor=red,
    filecolor=blue,   
    urlcolor=magenta} 

\renewcommand{\figurename}{FIGURE}
\renewcommand{\thefigure}{\arabic{figure}}
\def\bibsection{\refname}
\renewcommand{\refname}{\noindent\textbf{REFERENCES}}

\begin{document}
 
 \begin{center}
 
  \textbf{\large Observation of momentum-dependent charge density wave gap in a layered antiferromagnet  ${\mathbf{Gd}}{\mathbf{Te}}_3$}\\[.2cm]
  Sabin~Regmi$^{1,6}$, Iftakhar~Bin~Elius$^{1}$, Anup~Pradhan~Sakhya$^{1}$, Dylan~Jeff$^{1,2}$, Milo~Sprague$^{1}$, Mazharul~Islam~Mondal$^{1}$, Damani~Jarrett$^{1}$, Nathan~Valadez$^{1}$, Alexis~Agosto$^{1}$, Tetiana~Romanova$^{3}$, Jiun-Haw~Chu$^{4}$, Saiful~I.~Khondaker$^{1,2}$,  Andrzej~Ptok$^{5}$, Dariusz~Kaczorowski$^{3}$, Madhab~Neupane*$^{1}$\\[.2cm]
 {\itshape
    $^{1}$Department of Physics, University of Central Florida, Orlando, Florida  32816, USA\\
    $^{2}$NanoScience Technology Center, University of Central Florida,\\ Orlando, Florida 32826, USA\\
    $^{3}$Institute of Low Temperature and Structure Research, \\Polish Academy of Sciences, Ok\'{o}lna 2, PL-50-422 Wroc\l{}aw, Poland\\
    $^{4}$Department of Physics, University of Washington, Seattle, Washington 98195, USA\\
  	$^{5}$Institute of Nuclear Physics, Polish Academy of Sciences, \\W. E. Radzikowskiego 152, PL-31342 Krak\'{o}w, Poland\\
  	$^6$Present address: Center for Quantum Actinide Science and Technology, Idaho National laboratory, Idaho Falls, ID 83415, USA 	
  }
\\[.2cm]
$^*$Corresponding author: madhab.neupane@ucf.edu
\\[1cm]
\end{center}
\newpage

\textbf{Charge density wave (CDW) ordering has been an important topic of study for a long time owing to its connection with other exotic phases such as superconductivity and magnetism. The $R{\mathrm{Te}}_{3}$ ($R$ = rare-earth elements) family of materials provides a fertile ground to study the dynamics of CDW in van der Waals layered materials, and the presence of magnetism in these materials allows to explore the interplay among CDW and long range magnetic ordering. Here, we have carried out a high-resolution angle-resolved photoemission spectroscopy (ARPES) study of a CDW material ${\mathrm{Gd}}{\mathrm{Te}}_{3}$, which is antiferromagnetic below $\sim\mathrm{12~K}$, along with thermodynamic, electrical transport, magnetic, and Raman measurements. Our ARPES data show a two-fold symmetric Fermi surface with both gapped and ungapped regions indicative of the partial nesting. The gap is momentum dependent, maximum along $\mathrm{\overline{\Gamma}} - \mathrm{\overline{Z}}$ and gradually decreases going towards $\mathrm{\overline{\Gamma}} - \mathrm{\overline{X}}$. Our study provides a platform to study the dynamics of CDW and its interaction with other physical orders in two- and three-dimensions.}\\

Charge density wave (CDW) \cite{Gruner1988, Gruner1994} in quantum materials have been a subject of numerous research works for a number of decades because of its relevance in understanding several physical properties and also its competition or coexistence with exotic phases like superconductivity and magnetism \cite{Morris1975, Gabovich2000, Jung2000, Gabovich2001, Toyoka2001, Galli2002, Jung2003, Iyeiri2003, Singh2005, Fang2005, Hossain2005, Shimomura2009}. CDW ordering is a phenomenon associated with Fermi surface (FS) instabilities, where a periodic lattice distortion leads to the spatial modulation of carrier density \cite{Gruner1994}. One example is the Peierls distortion in one-dimension \cite{Peierls1955, Ashcroft1976}, in which the lattice periodicity can be doubled by electronically disturbing the system with a wave vector that is double the Fermi wave number, resulting into a gap opening at the Brillouin zone (BZ) boundaries nested by the same wave vector (FS nesting). With increase in dimensions, the FS nesting tends to be imperfect so that certain regions of the FS remain ungapped, leading to a metallic nature \cite{Gweon1998, Garcia2007}. The mechanism of CDW in such higher-dimensional materials is still of great interest as it can differ from material to material, can have different origin, and also depends on crystal growth conditions \cite{Zhu2017, Johannes2008, Thorne1992, Sayers2020}. \\

The orthorhombic crystalline family of van der Waals layered materials $R{\mathrm{Te}}_{3}$ ($R$~=~{rare-earth~elements}) has been broadly studied for the presence of CDW \cite{DiMasi1995, Laverock2005, Malliakas2005, Malliakas2006, Yumigeta2021}. The CDW ordering takes place at a high temperature, and materials with heavier rare-earth elements exhibit a second CDW transition at a lower temperature \cite{Ru2008a, Banerjee2013}. In addition, the existence of long range magnetic ordering in these compounds provides ground to study the interplay among magnetic and CDW orders \cite{Yuji2003, Ru2008b}. $R-\mathrm{Te}$ slabs are sandwiched in between planar bi-layers of $\mathrm{Te}-\mathrm{Te}$, where the neighboring $\mathrm{Te}-\mathrm{Te}$ are connected through weak van der Waals interaction, thereby easing the exfoliation of these layered materials to the two-dimensional limit \cite{YChen2019, Lei2020, Kogar2020, Watanabe2021}.  Angle-resolved photoemission spectroscopy (ARPES) \cite{Damascelli2003, Damascelli2004} has been a useful tool to directly probe the energy-momentum dispersion in quantum materials. It has been extensively used to study the electronic structure of $R{\mathrm{Te}}_{3}$ in the investigation of FS and CDW induced gap \cite{Gweon1998, Brouet2004, Komoda2004, Brouet2008, Schmitt2008, Moore2010, Lou2016, Lee2016, Seong2021, Liu2020, Chikina2023}. The gap size is found to depend on momentum, and the maximum gap changes as a function of the rare-earth element $R$, which can be modeled by a nesting driven sinusoidal CDW \cite{Brouet2008}.  Among the materials under this family, ${\mathrm{Gd}}{\mathrm{Te}}_{3}$ has been reported to have very high electronic mobility \cite{Lei2020} and steep band dispersion at the Fermi level \cite{Liu2020}. It can be thinned down to ultrathin limit using mechanical exfoliation that allows to study the thickness dependence of the CDW ordering  \cite{Lei2020, YChen2019}. Properties such as CDW ordering, magnetic ordering, and pressure induced superconductivity \cite{Ru2008a, Ru2008b, Zocco2015} make this system interesting in order to explore more on the electronic properties.  \\

In this article, by utilizing high-resolution ARPES measurements, we study the electronic structure of layered van der Waals material ${\mathrm{Gd}}{\mathrm{Te}}_{3}$. Our thermodynamic and electrical transport measurements show that the material is antiferromagnetic (AFM), with the magnetic transition occurring below $\sim \mathrm{12~K}$. The CDW transition occurs well above room temperature, at around $\sim\mathrm{375~K}$. Our ARPES results show two-fold symmetric FS with spectral intensity absent around certain regions of the FS, especially around $\mathrm{\overline{\Gamma}} - \mathrm{\overline{M}}$ to $\mathrm{\overline{\Gamma}} - \mathrm{\overline{Z}}$, indicating the presence of gap at the Fermi level. Some regions of the FS near the $\mathrm{\overline{X}}$ point remain gapless, implying partial nesting. This gap is strongly direction dependent, with a gap maximum along $\mathrm{\overline{\Gamma}} - \mathrm{\overline{Z}}$ and gradually decreasing  towards $\mathrm{\overline{\Gamma}} - \mathrm{\overline{X}}$. The room temperature Raman spectroscopy measurement shows the presence of CDW amplitude mode in the bulk as well as ultrathin samples up to 4L ($\sim\mathrm{5~nm}$). Our results indicate that this material is excellent for studying the dynamics and interaction of CDW with other physical parameters in both three- and two-dimensional limits.\\

High-quality single crystals of $\mathrm{Gd}\mathrm{Te}_3$ were synthesized using the self-flux technique, as described in the literature \cite{RuFisher2006}. Chemical composition and phase homogeneity of the crystals were determined by means of energy-dispersive x-ray (EDX) analysis performed using a FEI scanning electron microscope equipped with an EDAX Genesis XM4 spectrometer. The specimens examined were found homogeneous and single-phase. The crystal structure was verified by powder x-ray diffraction (XRD) made on finely pulverized crystals employing a PANanalytical X’pert Pro diffractometer with $\mathrm{Cu~K_\alpha}$ radiation. The XRD data was evaluated using the Rietveld method and the FULLPROF software package. The result confirmed the orthorhombic structure (space group \#63) and yielded the lattice parameters $a \approx c \approx \mathrm{4.33~\AA}$, $b = \mathrm{25.3~\AA}$. In addition, the single crystal selected for physical properties measurements was examined on an Oxford Diffraction X’calibur four-circle single-crystal x-ray diffractometer equipped with a CCD Atlas detector.

 Magnetic properties were investigated from $\mathrm{1.72~K}$ to $\mathrm{300~K}$ and in magnetic fields up to $\mathrm{7~T}$, applied perpendicular to the crystallographic $b$-axis, using a Quantum Design MPMS-XL superconducting quantum interference device (SQUID) magnetometer. The heat capacity was measured in the temperature interval $\mathrm{2-400~K}$ employing the relaxation technique and two-$\tau$ model implemented in a Quantum Design PPMS-9 platform. Electrical transport measurements were performed on a bar-shaped specimen cut from the oriented crystal using a wire saw. Electrical contacts were made by silver wires of diameter $\mathrm{20~\mu m}$, attached to the specimen's surface in a linear manner with a single-component silver paste. The experiments were carried out in the same PPMS platform in the temperature range $\mathrm{2-300~K}$ employing a standard four-points ac technique and electrical current flowing within the crystallographic $ac$-plane.

Raman spectroscopy measurements were performed in ambient conditions using a Horiba LabRAM HR Evolution Spectrometer equipped with a $\mathrm{1,800~grooves/mm}$ grating and a Synapse EMCCD detector. A frequency doubled Nd:YAG $\mathrm{532~nm}$ excitation laser source was focused to a square micron sized beam spot using an $\mathrm{100x}$ objective. An ultra-low-frequency (ULF) filter was utilized to resolve the CDW peak at low wavenumbers.

High-resolution ARPES measurements were performed at the Stanford Synchrotron Radiation Lightsource (SSRL) end-station 5-2 equipped with a DA30 electron analyzer. The angular and energy resolutions were set better than $\mathrm{0.2\degree}$ and $\mathrm{20~meV}$, respectively. The samples, mounted on copper posts and attached to ceramic posts on top using silver epoxy paste, were loaded and cleaved \textit{in situ} inside the ARPES chamber maintained under ultra high vacuum better than $\mathrm{10^{-10}~torr}$. Measurements were carried out at a temperature of $\mathrm{8~K}$. The Fermi level was determined by fitting the leading edge in the energy distribution curve of gold spectrum, and the gap was obtained by fitting the leading edge of the energy distribution curves (EDCs) and comparing the shift from the so-determined Fermi level.

The first-principles calculations based on density-functional theory (DFT) \cite{Hohenberg1964, Kohn1965} were performed using the projector augmented-wave (PAW) potentials~\cite{Blochl1994} implemented in the {\sc Quantum ESPRESSO }~\cite{Giannozzi2009, Giannozzi2017, Giannozzi2020}. The calculations were performed within the generalized gradient approximation (GGA) in the Perdew, Burke, and Ernzerhof (PBE) parameterization \cite{PBE1996}, develop within {\sc PsLibrary}~\cite{DalCorso2014}. The atom position was optimized for conventional unit cell with experimental values of lattice vectors, using the $\mathrm{14} \times \mathrm{2} \times \mathrm{14}$ {\bf k}--point grid in the Monkhorst--Pack scheme~\cite{Monkhorst1976}. As the convergence condition for the optimization loop, we took the energy difference of $\mathrm{10^{-6}~eV}$. The calculations were performed with the energy cut-off set to $\mathrm{400~eV}$. In calculations, the $\mathrm{Gd}~4f$ electrons were treated as valence states. To theoretical study of electronic band structure, we use the tight binding model in the maximally localized Wannier orbitals basis~\cite{Marzari1997, Souza2001}. This model was constructed from exact DFT calculations in a conventional unit cell, with $\mathrm{6} \times \mathrm{2} \times \mathrm{6}$ $\Gamma$-centered {\bf k}--point grid, using the {\sc Wannier90} software~\cite{Pizzi2020}. The tight binding ($\mathrm{28}$ orbitals $\mathrm{112}$ bands) model is based on $d$ orbitals of $\mathrm{Gd}$, and $p$ orbitals of $\mathrm{Te}$. Finally, the spectrum of the system were calculated using the surface Green’s function technique for a semi-infinite system \cite{MSancho1985}, implemented in WannierTools \cite{QWu2018}.

\begin{figure*} [ht!]
\includegraphics[width=1.0\textwidth]{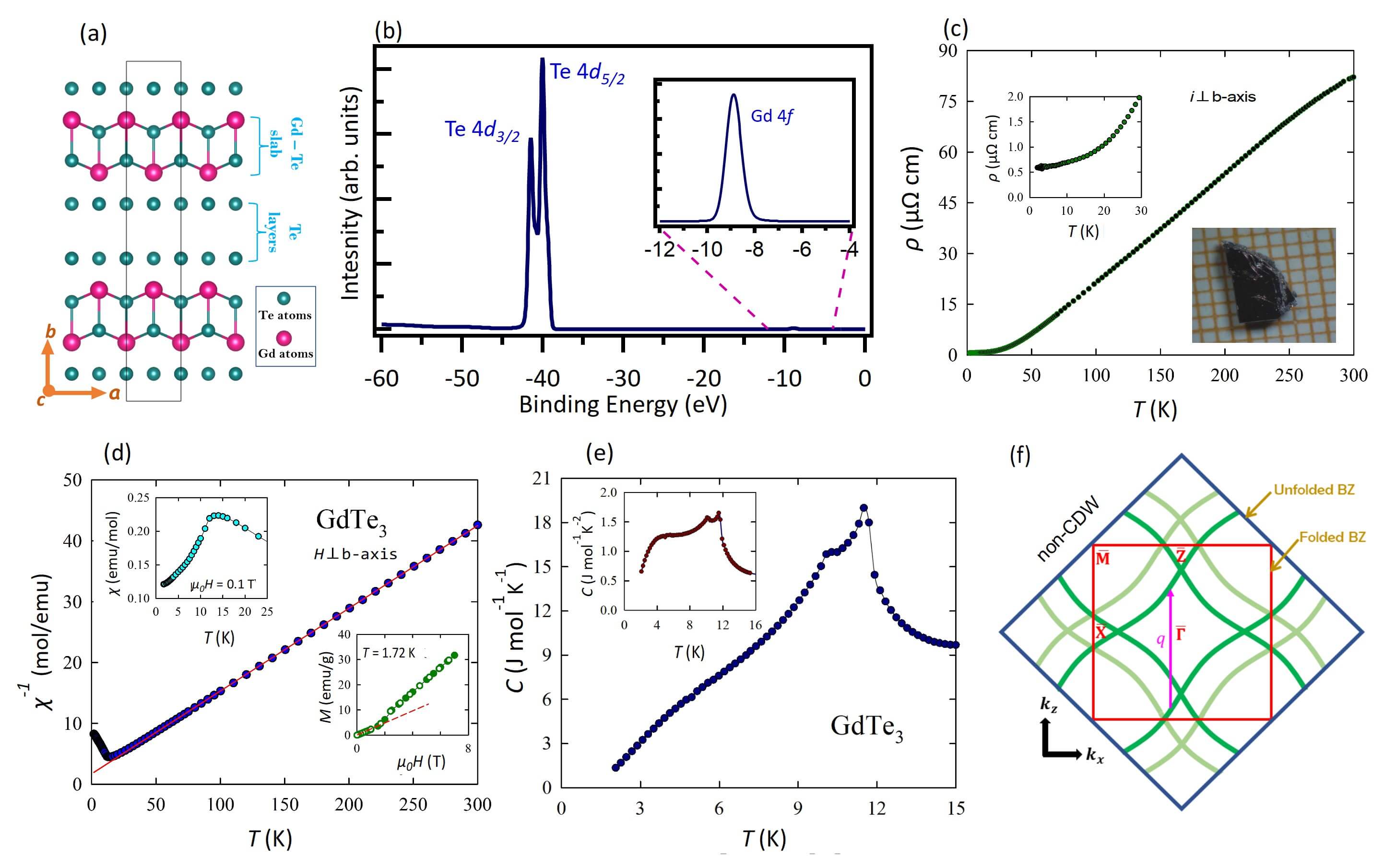}
\caption{Crystal structure and sample characterization of ${\mathrm{Gd}}{\mathrm{Te}}_{3}$. \textbf{a}~Crystal structure of ${\mathrm{Gd}}{\mathrm{Te}}_{3}$, where the pink and the teal colored balls represent $\mathrm{Gd}$ and $\mathrm{Te}$ atoms, respectively. \textbf{b}~Spectroscopic core level spectrum showing peaks of $\mathrm{Gd}~4f$ and $\mathrm{Te}~4d$ peaks. \textbf{c}~Temperature dependence of the electrical resistivity measured with electrical current flowing in the crystallographic $a-c$ plane. Inset: the low-temperature resistivity data. \textbf{d}~Temperature variation of the inverse magnetic susceptibility measured in a magnetic field of $\mathrm{0.1~T}$ applied perpendicular to the crystallographic $b$-axis. Solid line represents the Curie-Weiss fit described in the text. Top left inset: The low-temperature magnetic susceptibility data. Bottom right inset: Magnetic field variation of the magnetization taken at $\mathrm{1.72~K}$ with increasing (full circles) and decreasing (open circles) field strength. Dashed line emphasizes a linear behavior in small fields and the metamagnetic transition near $\mathrm{1.5~T}$. \textbf{e}~Low-temperature dependence of specific heat near the magnetic transition. Inset: Same data plotted as the ratio of specific heat to temperature versus temperature.  \textbf{f}~Schematic of non-CDW FS in $R\mathrm{Te}_3$ showing two-dimensional BZ of $\mathrm{Te}$ plane and three-dimensional BZ from the three-dimensional unit cell.}
\label{F1}
\end{figure*}

${\mathrm{Gd}}{\mathrm{Te}}_{3}$ crystallizes in a layered orthorhombic structure (space group Number 63) with lattice parameters $a \approx c \approx \mathrm{4.33~{\AA}}$ and $b = \mathrm{25.28~\AA}$, close to the values reported in literature \cite{Lei2020}. In Figure \ref{F1}a, we present the side view of the crystal structure of ${\mathrm{Gd}}{\mathrm{Te}}_{3}$, where pink balls represent the gadolinium atoms and teal colored balls represent the tellurium atoms. The crystal structure is composed of $\mathrm{Gd - Te}$ slabs sandwiched in between the $\mathrm{Te}$ bi-layers. Neighboring $\mathrm{Te}$ layers are bonded by weak van der Waals interaction, and the natural cleaving plane is parallel to the \textit{ac} plane - the (010) plane. In Figure \ref{F1}b, we show the  spectroscopic core level spectrum, where peaks associated with $\mathrm{Te}$ 4$d$ and $\mathrm{Gd}$ 4$f$ can be clearly identified. 

${\mathrm{Gd}}{\mathrm{Te}}_{3}$ exhibits a CDW phase transition, with the transition temperature  well above room temperature ($\sim \mathrm{377~K}$) \cite{Ru2008a}. In addition, it also undergoes an AFM transition at a low temperature of $\sim\mathrm{12~K}$ \cite{Ru2008b, Lei2020, Guo2021}. Our heat capacity ($C$) measurements also show an anomaly near $\mathrm{375~K}$ (see Supplementary Note 1 \& Supplementary Figure 1] that arises due to the incommensurate charge density wave formation \cite{Yumigeta2021}. The room temperature Raman spectrum for the bulk crystal shows a CDW amplitude mode along with other phonon modes, in concert with the literature \cite{Wang2022, YChen2019}, establishing the presence of room-temperature CDW in the material. The mode remains prominent when the bulk crystal is thinned down to 4-layered samples via gold-assisted mechanical exfoliation, which suggests this material to be an excellent platform to study the interplay of CDW and long-range orders down to the two-dimensional limit as well [see Supplementary Note 2 \& Supplementary Figure 2]. Figure~\ref{F1}c displays the temperature dependence of the electrical resistivity ($\rho$) of ${\mathrm{Gd}}{\mathrm{Te}}_{3}$ measured with electrical current flowing in the crystallographic $ac$-plane. In concert with the previous reports \cite{Ru2008a, Ru2008b, Lei2020}, single crystals of ${\mathrm{Gd}}{\mathrm{Te}}_{3}$ investigated in the present study exhibit very good metallic-type charge conductance. The ratio of the resistivity values measured at $\mathrm{300~K}$ and $\mathrm{2~K}$ is as large as $\mathrm{140}$, thus indicating high crystalline quality of the samples. On approaching the room temperature, the $\rho(T)$ curve clearly changes its slope signaling the proximity of the CDW transition. Figure~\ref{F1}d displays the temperature dependence of the inverse magnetic susceptibility ($\chi^{-1}$) measured for magnetic field applied perpendicular to the crystallographic $b$-axis. Above about $\mathrm{30~K}$, $\chi^{-1}(T)$ exhibits a straight line behavior that can be approximated by the Curie-Weiss formula with the effective magnetic moment $\mu_{eff} = \mathrm{7.67~\mu_B}$ and the paramagnetic Curie temperature $\theta_p = \mathrm{-12.6~K}$. The value of $\mu_{eff}$ is close to the theoretical prediction for $\mathrm{Gd}^{3+}$ ion. The negative value of $\theta_p$ reflects the predominance of AFM exchange interactions, which bring about the long range AFM ordering below about $\mathrm{12~K}$ (see the upper inset to Figure~\ref{F1}d). The AFM nature of the electronic ground state in ${\mathrm{Gd}}{\mathrm{Te}}_{3}$ is further corroborated by the characteristic behavior of the magnetization isotherm taken at $T = \mathrm{1.72~K}$ (see the lower inset in Figure~\ref{F1}d) with a clear metamagnetic transition near $\mathrm{1.5~T}$. Overall, the magnetic data collected in our study agrees very well with those reported in the literature \cite{Lei2020, Yuji2003, Guo2021}. We also observed a distinct lambda-shaped anomaly at around $\mathrm{11.6~K}$ in $C(T)$ graph, followed by a subsequent feature observed at $\mathrm{10~K}$ (Figure~\ref{F1}e), similar to previous reports \cite{Ru2008b, Lei2020, Guo2021}. \\

In Figure~\ref{F1}e, we present a schematic of the non-CDW FS of $R\mathrm{Te}_3$. The bands crossing the Fermi level come from the $p$ orbitals of the atoms within the $\mathrm{Te}$ layers. The two-dimensional unfolded band structure (dark green bands referred to as main bands hereafter) corresponds to "true" square lattice with one $\mathrm{Te}$ atom in the primitive unit cell. Such unit cell can be obtained by $\sqrt{2}$ times reduction and $45\degree$ rotation of the unit cell of the whole crystal structure. To account for the three-dimensional lattice symmetry, the FS is acquired by considering the folding of the bands in the $\mathrm{Te}$ plane leading to the folded BZ represented by the red square in Figure~\ref{F1}e.  The band folding is reflected on the observed band structure, where the mismatch between bands lead to their occurrence as the low intensity shadow bands~\cite{Brouet2008, Moore2010}. In Figure~\ref{F1}e, $q = \frac{5}{7}c^*$ represents a nesting condition, where the wave vector nests two sets of the main bands, which would still be present if we only consider the two-dimensional unit cell of the $\mathrm{Te}$ plane without folding.

\begin{figure*} [ht!]
\includegraphics[width=1\textwidth]{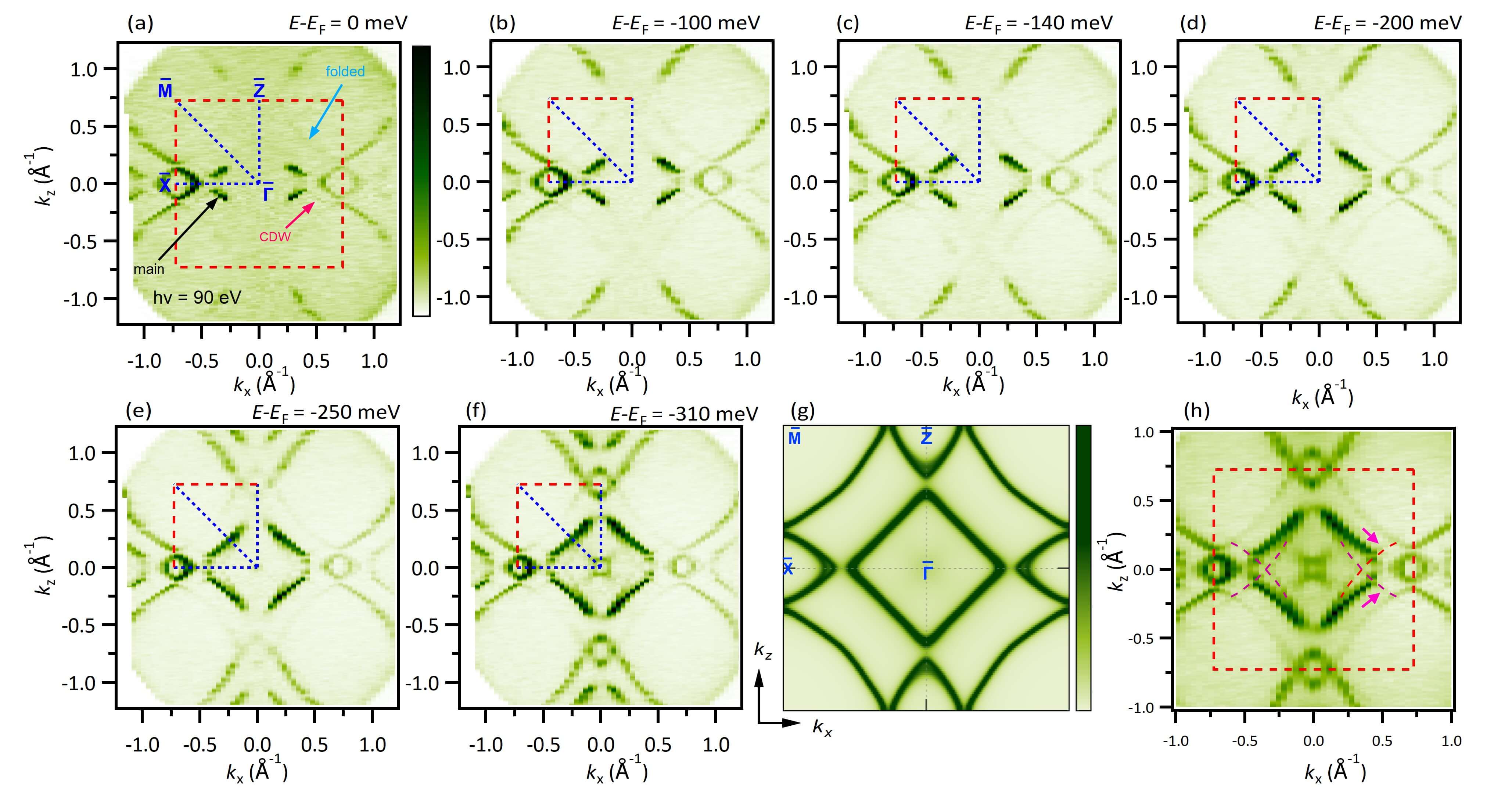}
\caption{Constant energy contours on the (010) surface of ${\mathrm{Gd}}{\mathrm{Te}}_{3}$. \textbf{a}~ARPES measured FS with a photon energy of $\mathrm{90~eV}$. Experimental BZ is shown with red dashed lines. High-symmetry points and directions are labeled in blue color. \textbf{b-f}~Energy contours at various binding energies as noted on top of each plot. \textbf{g}~Calculated FS for non-CDW case. \textbf{h}~Energy contour integrated within $\mathrm{50~meV }$ window centered at $\mathrm{300~meV}$ binding energy. ARPES data were collected at the SSRL end-station 5-2 at a temperature of $\mathrm{8~K}$. }
\label{F2}
\end{figure*}

The constant energy contours obtained at the FS and at various binding energies using a photon source of energy $\mathrm{90~eV}$ ($T = \mathrm{8~K}$) are presented in Figure \ref{F2}. The BZ represented by the dashed red lines is obtained from the (010) surface projection of the three-dimensional BZ. As seen in Figure \ref{F2}a, the FS is metallic, in agreement with the metallic nature observed in the transport measurements, and is two-fold symmetric. Strong photoemission intensity is observed at the Fermi level along and around the $\mathrm{\overline{\Gamma}} - \mathrm{\overline{X}}$ direction. However, the intensity for the main bands along the  $\mathrm{\overline{\Gamma}} - \mathrm{\overline{M}}$ and $\mathrm{\overline{\Gamma}} - \mathrm{\overline{Z}}$ directions is missing.  On going to lower binding energy of about $\mathrm{100~meV}$, the intensity of the main bands starts to fill up towards the $\mathrm{\overline{\Gamma}} - \mathrm{\overline{M}}$ line, however, a gap still exists along the  $\mathrm{\overline{\Gamma}} - \mathrm{\overline{M}}$ and $\mathrm{\overline{\Gamma}} - \mathrm{\overline{Z}}$ directions. The main band intensity only fills up along the  $\mathrm{\overline{\Gamma}} - \mathrm{\overline{M}}$ line at around $\sim \mathrm{140~meV}$ binding energy and along the  $\mathrm{\overline{\Gamma}} - \mathrm{\overline{Z}}$ line at $\sim\mathrm{310~meV}$ binding energy. These results indicate that a gap exists at the Fermi level along the $\mathrm{\overline{\Gamma}} - \mathrm{\overline{M}}$ and $\mathrm{\overline{\Gamma}} - \mathrm{\overline{Z}}$ directions. Similar nature of the gap was obtained in different set of measurements performed using  $\mathrm{68~eV}$ incident photon energy, in which the main band intensity appears along the $\mathrm{\overline{\Gamma}} - \mathrm{\overline{M}}$ and $\mathrm{\overline{\Gamma}} - \mathrm{\overline{Z}}$  directions, only around the binding energies of $\sim\mathrm{150~meV}$ and $\sim \mathrm{320~meV}$ binding energies, respectively [see Supplementary Note 3 \& Supplementary Figure 3]. In Figure~\ref{F2}h, we present the theoretically calculated FS (without considering CDW), which is similar to the one presented in Figure~\ref{F1}e. To compare with the theoretical FS, we present the experimental energy contour in Figure~\ref{F1}e, which is integrated up to $\sim\mathrm{300~meV}$ binding energy.  Main bands as well as  the shadow bands (from the folding), as previously described, can be observed. In addition to the low intensity bands coming from the folding of the three-dimensional BZ, we also observe other faint bands (shown by magenta colored arrows and traced by magenta colored curves) that are not captured in the calculations [Also see Supplementary Note 4 and Supplementary Figure 4]. These bands arise as a result of the CDW ordering.\\
 
\begin{figure*} [ht!]
\includegraphics[width=1\textwidth]{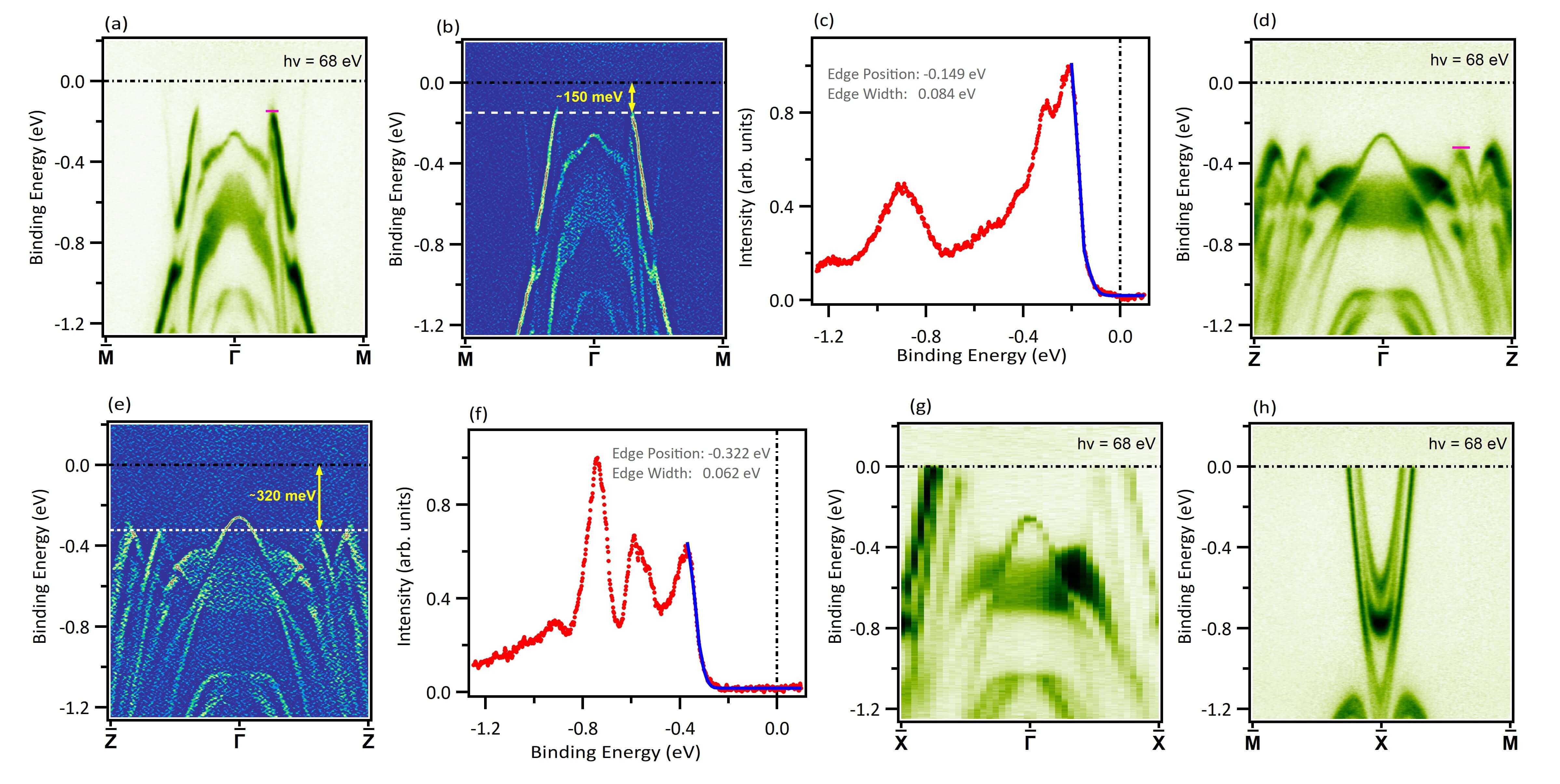}
\caption{Experimental observation of CDW induced gap. \textbf{a}~Dispersion map and \textbf{b}~its second derivative along the $\mathrm{\overline{M}}-\mathrm{\overline{\Gamma}} - \mathrm{\overline{M}}$ direction. \textbf{c}~Energy distribution curve integrated within the momentum window represented by magenta line in (a) and the Fermi edge fit. \textbf{d}~Dispersion map and \textbf{e}~its second derivative along the  $\mathrm{\overline{Z}}-\mathrm{\overline{\Gamma}} - \mathrm{\overline{Z}}$ direction. \textbf{f}~Energy distribution curve integrated within the momentum window represented by the magenta line in (d) and the Fermi edge fit. \textbf{g} Experimental dispersion map along the $\mathrm{\overline{X}} - \mathrm{\overline{\Gamma}} - \mathrm{\overline{X}}$ direction. \textbf{h}~Experimental band structure along $\mathrm{\overline{M}} - \mathrm{\overline{X}} - \mathrm{\overline{M}}$. Data were collected at the SSRL beamline 5-2 at a temperature of $\mathrm{8~K}$.}
\label{F3}
\end{figure*}

In order to quantify the gap along the $\mathrm{\overline{\Gamma}} - \mathrm{\overline{M}}$ and $\mathrm{\overline{\Gamma}} - \mathrm{\overline{Z}}$ directions, we took cuts along these directions and analyzed the corresponding band dispersion. In Figure \ref{F3}a, we present the dispersion map along $\mathrm{\overline{\Gamma}} - \mathrm{\overline{M}}$ obtained using a photon source of $\mathrm{68~eV}$ at a temperature of $\mathrm{8~K}$. From the band dispersion, there exists a gap along this direction line with the absence of photoemission signal in the FS. The second derivative plot of the band dispersion presented in Figure \ref{F3}b shows that the bands extend up to the binding energy of about $\sim\mathrm{150~meV}$, and a clear gap exists above this binding energy along the $\mathrm{\overline{\Gamma}} - \mathrm{\overline{M}}$ direction. The gap of  about $\mathrm{150~meV}$ below the Fermi level can also be seen from the Fermi fit of the leading edge in the energy distribution curve (Figure~\ref{F3}c) taken within the momentum window represented by the magenta colored solid  line in Figure~\ref{F3}a. Next, we turn our attention to explore the gap along $\mathrm{\overline{\Gamma}} - \mathrm{\overline{Z}}$. From the dispersion map (Figure~\ref{F3}d), its second derivative (Figure \ref{F3}e), and the Fermi fit of the leading edge in energy distribution curve (Figure~\ref{F3}f), it is clear that a gap below the Fermi level of about $\sim\mathrm{320~meV}$ exists along this direction. The dispersion maps for $\mathrm{90~eV}$ incident photon energy are presented in the Supplementary Figure~5 [also see Supplementary Note 5], which show that gaps along the $\mathrm{\overline{\Gamma}} - \mathrm{\overline{M}}$ and $\mathrm{\overline{\Gamma}} - \mathrm{\overline{Z}}$ directions are $\sim \mathrm{140~meV}$ and $\sim \mathrm{310~meV}$, respectively, as observed in the energy contours presented in Figure~\ref{F2}. The calculated band structures along these directions well reproduce the experimental data barring the CDW induced gap as the calculations are carried out for the non-CDW case. ARPES can only probe up to the Fermi energy, so the value of the total gap size can not be obtained from the ARPES measurements. Along $\mathrm{\overline{\Gamma}} - \mathrm{\overline{X}}$ and $\mathrm{\overline{M}} - \mathrm{\overline{X}}$, however, bands cross the Fermi level [Figure~\ref{F3}g,h; also see Supplementary Note 6 \& supplementary Figure~6], which is in accordance with the photoemission intensity observed along this direction in the energy contours in Figure~\ref{F2}. \\

\begin{figure*} [ht!]
\includegraphics[width=1\textwidth]{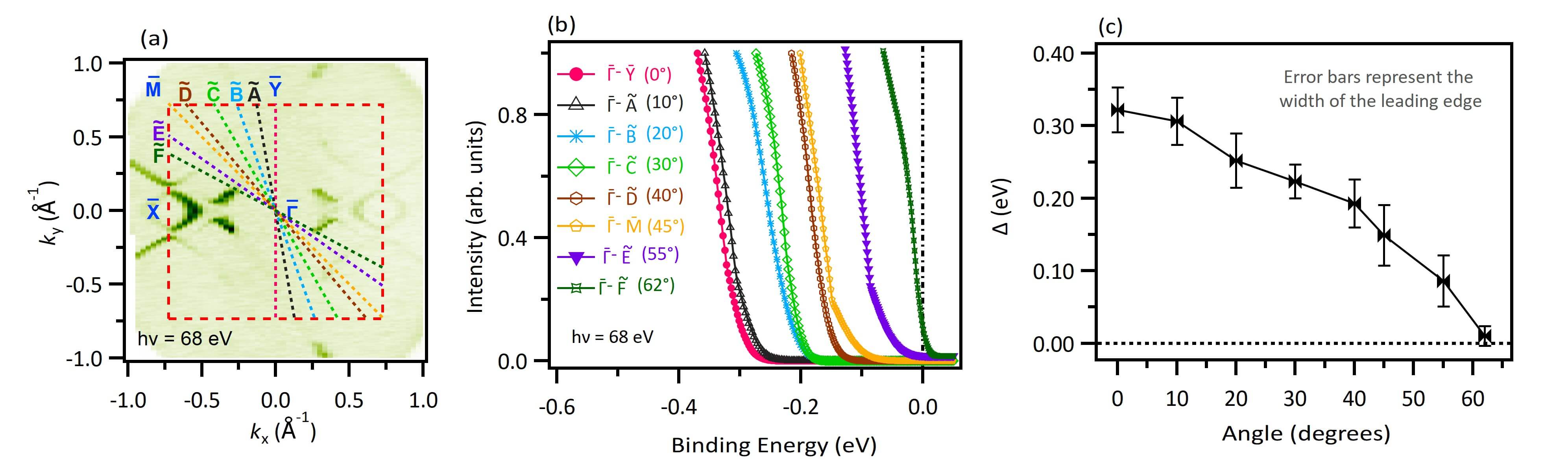}
\caption{Momentum dependence of the gap. \textbf{a}~FS measured with photon energy of $\mathrm{68~eV}$ with colored lines representing the cuts at color-coded angles. The angles are taken with respect to the $\mathrm{\overline{Z}}-\mathrm{\overline{\Gamma}} - \mathrm{\overline{Z}}$ direction. \textbf{b}~Shift in the fitted Fermi edge on going away from $\mathrm{\overline{Z}}-\mathrm{\overline{\Gamma}} - \mathrm{\overline{Z}}$. \textbf{c}~Gap below the Fermi level plotted against the angle from the $\mathrm{\overline{Z}}-\mathrm{\overline{\Gamma}} - \mathrm{\overline{Z}}$ direction. Data were collected at the SSRL beamline 5-2 at a temperature of $\mathrm{8~K}$.}
\label{F4}
\end{figure*}

From the observations in Figures~\ref{F2} and \ref{F3}, we get the idea that  the spectral intensity corresponding to the main bands appears at lower and lower binding energies as we move from $\mathrm{\overline{\Gamma}} - \mathrm{\overline{X}}$ to $\mathrm{\overline{\Gamma}} - \mathrm{\overline{Z}}$. In Figure~\ref{F4}, we analyze this momentum dependence of the gap below the Fermi level in measurements using a photon energy of $\mathrm{68~eV}$ as a function of counter-clockwise angle from the $\mathrm{\overline{\Gamma}} - \mathrm{\overline{Z}}$ direction. $\mathrm{0\degree}$ corresponds to the $\mathrm{\overline{\Gamma}} - \mathrm{\overline{Z}}$ direction and a $\mathrm{45\degree}$ counter-clockwise rotation means we are looking at the $\mathrm{\overline{\Gamma}} - \mathrm{\overline{M}}$ direction. Figure~\ref{F4}b represents the leading edges in the integrated energy distribution curves for dispersion maps corresponding to different angles defined in Figure~\ref{F4}a. It is clear that the leading edge shifts towards the Fermi level when we take the dispersion map away from the $\mathrm{\overline{\Gamma}} - \mathrm{\overline{Z}}$ direction towards the $\mathrm{\overline{\Gamma}} - \mathrm{\overline{X}}$ direction, indicative of reducing gap below the Fermi level. At an angle of $\mathrm{62~\degree}$, the gap size surpasses the limits of the experimental energy resolution, therefore, it can be anticipated that the gap vanishes at greater angles [also see Supplementary Note 7 \& Supplementary Figure 7]. In Figure \ref{F4}c, we plot the gap below the Fermi level as a function of the counter-clockwise angle from $\mathrm{\overline{\Gamma}} - \mathrm{\overline{Z}}$, where with angle, the gap gradually reduces from around $\mathrm{320~meV}$ at $\mathrm{0\degree}$ to around $\mathrm{150~meV}$ at $\mathrm{45\degree}$, i.e., $\mathrm{\overline{\Gamma}} - \mathrm{\overline{M}}$ direction and reaches out of experimental resolution at angles greater than $\mathrm{62\degree}$. \\

Given the existence of CDW ordering and AFM ordering in $R\mathrm{Te}_3$ materials, in which $4f$ electrons coming from the $R$-element may bring electronic correlations into play, these materials are excellent platforms to study the interplay among electronic interactions, CDW,  and long-range magnetic orders. The layered nature of the crystal structure with very weak van der Waals interaction provides opportunity to study such interplay in both three- and two-dimensions as mechanically exfoliating the crystals to the two-dimensional limit is easy. In this study, we studied bulk crystals of one of such materials $\mathrm{Gd}\mathrm{Te}_3$, which exhibits CDW transition at $\sim\mathrm{375~K}$ and the magnetic ordering onsets below $\sim \mathrm{12~K}$ \cite{Ru2008a, Ru2008b, Iyeiri2003, Lei2020, Guo2021}. The $\mathrm{Gd}~4f$ states are well below the Fermi level as seen in the spectroscopic core level measurement.  We were able to obtain thin flakes of $\mathrm{Gd}\mathrm{Te}_3$ via mechanical exfoliation, which still showed prominent CDW amplitude peak in the Raman spectrum taken at room temperature, indicating the presence of high-temperature CDW in the ultrathin limit as well. In fact, the CDW transition temperature is expected to increase with lowering of the sample thickness \cite{YChen2019}. The FS obtained from ARPES measurement is two-fold symmetric. This two-fold symmetry instead of four-fold symmetry is as a result of slightly greater $c$ than $a$ ($c-a\sim \mathrm{0.01~\AA}$ \cite{Lei2020}), and this changes possible two degenerate CDWs along $a$ and $c$ axes in favor of a single CDW along $c$ direction with a CDW wave-vector $q_{CDW} = \frac{2}{7}c^*$ \cite{DiMasi1995, Malliakas2006, Ru2008a}. This wave vector connects the inner and outer diamond sheets in the FS, formed by the unfolded and folded band structures, respectively. An equivalent wave vector $q = c^*-q_{CDW} = \frac{5}{7}c^*$ nests the FS made by the main bands coming from a "true" two-dimensional plane of $\mathrm{Te}$ atoms and therefore would be present even if the effects of the band folding scenario are not considered \cite{Fang2007, Brouet2004, Ru2008a}. In our experimental data, the FS features coming from the shadow folded bands seem to be present even at the FS, however, the main band intensity is only along and around the $\mathrm{\overline{\Gamma}} - \mathrm{\overline{X}}$ direction and absent going away from this direction around $\mathrm{\overline{\Gamma}} - \mathrm{\overline{M}}$ and $\mathrm{\overline{\Gamma}} - \mathrm{\overline{Z}}$ directions. Therefore, our study shows that the CDW gap in $\mathrm{Gd}\mathrm{Te}_3$ is governed by the nesting condition $q = \frac{5}{7}b^*$. Probing of the CDW gap is restricted to below the Fermi level, as the ARPES spectral function depends upon Fermi-Dirac function, which is zero above the Fermi energy. The gap below the Fermi level is strongly dependent on the direction along which we take the dispersion map - it is highest along $\mathrm{\overline{\Gamma}} - \mathrm{\overline{Z}}$ and reduces gradually towards $\mathrm{\overline{\Gamma}} - \mathrm{\overline{X}}$, vanishing before reaching $\mathrm{\overline{\Gamma}} - \mathrm{\overline{X}}$ line. Such a nature of the gap occurs in nesting driven CDW when the nesting is imperfect and has been reported in a previous study on $R{\mathrm{Te}}_{3}$  which reports the nesting to become imperfect away from  $k_x = 0$ \cite{Brouet2008}. Measurement performed at a different temperature of $\mathrm{15~K}$ (conducted at a different beamline setup) reveals similar dependence on momentum (see Supplementary~Figure~8). In addition to the CDW induced gap, we are able to identify some features that can not be described in the non-CDW FS with main and folded bands. These features also appear as shadow-like bands with weak intensity and are expected to originate due to the reconstruction caused by the CDW ordering. \\

To conclude, we studied a van der Waals layered material, ${\mathrm{Gd}}{\mathrm{Te}}_{3}$, by performing high-resolution ARPES measurements of the electronic structure. We were able to probe the CDW induced gap as well as band features originating from the CDW ordering-induced reconstruction.  The gap obtained in our study is strongly dependent on the momentum direction with the highest gap lying along the $\mathrm{\overline{\Gamma}} - \mathrm{\overline{Z}}$ direction. A prominent peak associated with the CDW amplitude mode is seen  in our Raman measurements in samples as thin as 4L. Overall, our study demonstrates that ${\mathrm{Gd}}{\mathrm{Te}}_{3}$ is an excellent material platform to investigate the dynamics of CDW and its connection with other physical orders like magnetism and superconductivity as a function of sample thickness ranging from three- to two-dimensions.\\

\textit{Acknowledgments -} M.N. acknowledges the support from the National Science Foundation (NSF) CAREER award DMR-1847962,   the NSF Partnerships for Research and Education in Materials (PREM) Grant DMR-2121953, and the Air Force Office of Scientific Research MURI Grant No. FA9550-20-1-0322. D.K. was supported by the National Science Centre (Poland) under research grant 2021/41/B/ST3/01141. A.P. acknowledges the support by National Science Centre (NCN, Poland) under Projects No. 2021/43/B/ST3/02166 and also appreciates the funding in the frame of scholarships of the Minister of Science and Higher Education (Poland) for outstanding young scientists (2019 edition, No. 818/STYP/14/2019). S.I.K. acknowledges the support from the NSF PREM Grant DMR-2121953. The use of SSRL in SLAC National Accelerator Laboratory is supported by the U.S. Department of Energy (DOE), Office of Science, Office of Basic Energy Sciences under Contract No. DE-AC02-76SF00515. This research also used the resources of the Advanced Light Source (ALS), which is a DOE Office of Science User Facility under contract no. DE-AC02-05CH11231. We are grateful to Makoto Hashimoto \& Donghui Lu at SSRL and Alexei Fedorov \& Sung-Kwan Mo at ALS for the beamline assistance.\\

\clearpage

\setcounter{figure}{0}
\renewcommand{\figurename}{\textbf{Supplementary Figure}}
\renewcommand{\thefigure}{{\textbf{\arabic{figure}}}}
\renewcommand{\tablename}{Supplementary Table}
\renewcommand{\thetable}{\arabic{table}}
\def\bibsection{\refname}
\renewcommand{\refname}{\noindent\textbf{Supplementary References}}
\begin{center}
\textbf{\Large SUPPLEMENTARY INFORMATION}
\end{center}

\section*{\uppercase{Supplementary Note 1: Sample Characterization and Heat capacity measurements}}
High-quality single crystals of $\mathrm{Gd}\mathrm{Te}_3$ were synthesized using the self-flux technique. Chemical composition and phase homogeneity of the crystals were determined by means of energy-dispersive X-ray (EDX) analysis performed using a FEI scanning electron microscope equipped with an EDAX Genesis XM4 spectrometer. The result of EDX analysis is presented in Supplementary~Figure~\ref{FigS1}a, which indicates the desired stoichiometry and homogeneous chemical composition with 26.27 \% $\mathrm{Gd}$ and 73.73 \% $\mathrm{Te}$.  Supplementary~Figure~\ref{FigS1}b shows the specific heat ($C$) of  $\mathrm{Gd}\mathrm{Te}_3$ measured up to $\mathrm{400~K}$, with inset showing a zoomed in view at high temperatures. Despite severe irregularities in the $C(T)$ curve caused by addenda contribution (Apiezon N grease), one can recognize an anomaly near $\mathrm{375~K}$ that arises due to the incommensurate charge density wave formation \cite{Yumigeta2021S}.  

\section*{\uppercase{Supplementary Note 2: Raman spectroscopy and mechanical exfoliation of thin flakes}}
The exfoliation of the thin layer $\mathrm{Gd}\mathrm{Te}_3$ samples was performed via $\mathrm{Au}$-Assisted exfoliation. The metal layer deposition was completed in an electron/thermal evaporation system (Thermionics E-beam \& RDM Thermal evaporator). A $\mathrm{2~nm}$ metal adhesion layer of $\mathrm{Ti}$ was first evaporated onto a $\mathrm{Si/SiO_2}$ substrate via e-beam evaporation. $\mathrm{5~nm}$ of $\mathrm{Au}$ was then evaporated onto the $\mathrm{Ti}$ adhesion layer via thermal evaporation. Soon after exposing the $\mathrm{Au}$ layer to ambiance, the $\mathrm{Gd}\mathrm{Te}_3$ crystals were cleaved from a parent crystal and lightly pressed onto the target substrate using Nitto tape (SPV-224 PVC) for one minute. Lifting the tape resulted in exfoliated thin layers of $\mathrm{Gd}\mathrm{Te}_3$ on the $\mathrm{Au/Ti/SiO_2/Si}$ substrate.

Atomic force microscopy (AFM) was utilized to measure the thickness and identify the layer number of the exfoliated thin layer $\mathrm{Gd}\mathrm{Te}_3$. AFM measurements were taken in non-contact mode using a SmartSPM1000 scanning probe microscope. A single layer of $\mathrm{Gd}\mathrm{Te}_3$ is reported to have a height of $\mathrm{1.2~nm}$, serving as the baseline for the measurement of the flake’s layer number. The AFM measurements were taken in ambient conditions. Optical image of representative thin flakes used in the Raman measurements are presented in Supplementary~Figure~\ref{FigS2}a, where the arrow represents the location of where the Raman was measured. The AFM images are presented in Supplementary~Figure~\ref{FigS2}b, with red line indicating the direction which height profile was taken. The height profile of corresponding thin flakes are presented in Supplementary~Figure~\ref{FigS2}c.

In order to ensure that the Raman spectra being reported was of pristine conditions, the bulk $\mathrm{Gd}\mathrm{Te}_3$ was cleaved before any Raman experimentation was done. The Raman spectra for bulk as well as for different layered samples up to 4L thin are presented in Supplementary~Figure~\ref{FigS2}d. In the Raman spectrum, several peaks are observed, which is consistent with those reported in the literature \cite{Wang2022S, YChen2019S}. The peaks at $\mathrm{64~cm^{-1}}$ and $\mathrm{79~cm^{-1}}$ represent overlapping degenerate $\mathrm{A_g}$ and $\mathrm{B_g}$ phonon modes, the peak at $\mathrm{59~cm^{-1}}$ is a two-fold symmetric phonon mode, the peaks at $\mathrm{104~cm^{-1}}$, $\mathrm{117~cm^{-1}}$, $\mathrm{129~cm^{-1}}$, $\mathrm{144~cm^{-1}}$ are $\mathrm{B_g}$ symmetric phonons, and the peak at $\mathrm{91~cm^{-1}}$ is an $\mathrm{A_g}$ symmetric phonon \cite{Wang2022S}. Of interest is the peak at ~$\mathrm{46~cm^{-1}}$, which has been reported to be the CDW amplitude mode \cite{Wang2022S, YChen2019S}. The presence of this strong peak at room temperature Raman data supports the presence of CDW phase at room temperature. In thin films up to 4L, this peak remains strong, indicative of the presence of room-temperature CDW in two-dimensional limit. In fact, it has been documented that the CDW transition shifts to higher temperature with diminishing thickness \cite{YChen2019S}.  Thus, $\mathrm{Gd}\mathrm{Te}_3$ could be an excellent platform for studying how different physical orders such as long-range magnetic, superconducting, charge-density wave, etc. in both three- and two-dimensions.

\section*{\uppercase{Supplementary Note 3: ARPES measured constant energy contours with photon energy of 68 \lowercase{e}V}}
In Supplementary~Figure~\ref{FigS3}, we present the Fermi surface and constant energy contours measured with a photon energy of $\mathrm{68~eV}$ obtained at $\mathrm{8~K}$. Similar to the observations in the main text Figure 2 (with $\mathrm{90~eV}$ photon energy), the Fermi surface has some intensity along and around the $\mathrm{\overline{X}}-\mathrm{\overline{\Gamma}}-\mathrm{\overline{X}}$ direction and no intensity along $\mathrm{\overline{M}}-\mathrm{\overline{\Gamma}}-\mathrm{\overline{M}}$ and $\mathrm{\overline{Z}}-\mathrm{\overline{\Gamma}}-\mathrm{\overline{Z}}$. Spectral intensity appears at $\sim\mathrm{150~meV}$ binding energy along $\mathrm{\overline{M}}-\mathrm{\overline{\Gamma}}-\mathrm{\overline{M}}$ and about $\sim\mathrm{320~meV}$ binding energy along  $\mathrm{\overline{Z}}-\mathrm{\overline{\Gamma}}-\mathrm{\overline{Z}}$.\\

\section*{\uppercase{Supplementary Note 4: Observation of CDW bands}}
In Supplementary~Figure~\ref{FigS4}a, we present an integrated energy contour taken with $\mathrm{68~eV}$ incident photon energy. The integration is done within a $\mathrm{50~meV}$ window centered at the binding energy of $\mathrm{300~meV}$. Similar to the observation for $\mathrm{90~eV}$ photon energy presented in the main text Figure~2h, the features that can not be described by the non-CDW calculations appear (shown by magenta colored arrow in \ref{FigS4}a and represented by magneta colored dashed curves in \ref{FigS4}b). An energy-momentum dispersion cut taken along the yellow colored dashed line in \ref{FigS4}b  at $k_x=\mathrm{-0.5~\AA^{-1}}$ is presented in Supplemental Figure~\ref{FigS4}c, where the intense main bands and shadow bands coming from both the band folding and CDW can be observed.

\section*{\uppercase{Supplementary Note 5: $\overline{\mathrm{M}}-\overline{\mathrm{\Gamma}}-\overline{\mathrm{M}}$ and $\overline{\mathrm{Z}}-\overline{\mathrm{\Gamma}}-\overline{\mathrm{Z}}$ dispersion maps  (90~\lowercase{e}V)}}
In Supplementary~Figure~\ref{FigS5}, we present the dispersion maps taken along the $\mathrm{\overline{M}}-\mathrm{\overline{\Gamma}}-\mathrm{\overline{M}}$ and $\mathrm{\overline{Z}}-\mathrm{\overline{\Gamma}}-\mathrm{\overline{Z}}$ directions taken from measurements using $\mathrm{90~eV}$  photon energy. From the dispersion maps (\ref{FigS5}a and \ref{FigS5}d) and their second derivative plots (\ref{FigS5}b and \ref{FigS5}e), it is clear that the gap size below the Fermi level for $\mathrm{90~eV}$ data are around $\mathrm{140~meV}$ and $\mathrm{310~meV}$ along $\mathrm{\overline{M}}-\mathrm{\overline{\Gamma}}-\mathrm{\overline{M}}$ and $\mathrm{\overline{Z}}-\mathrm{\overline{\Gamma}}-\mathrm{\overline{Z}}$, respectively in agreement with the observations in the constant energy contours presented in  the main text Figure~2. In Supplementary~Figures~\ref{FigS5}c and \ref{FigS5}f, we present the theoretically calculated surface spectrum along the $\mathrm{\overline{M}}-\mathrm{\overline{\Gamma}}-\mathrm{\overline{M}}$ and $\mathrm{\overline{Z}}-\mathrm{\overline{\Gamma}}-\mathrm{\overline{Z}}$ directions, respectively. While the spectrum overall matches the experimental observation, the gap is not seen in the calculations as they are carried out for the non-CDW case. \\

\section*{\uppercase{Supplementary Note 6: Dispersion maps along $\mathrm{\overline{X}}-\mathrm{\overline{\Gamma}}-\mathrm{\overline{X}}$ and $\mathrm{\overline{M}}-\mathrm{\overline{X}} - \mathrm{\overline{M}}$ }}
Supplementary~Figure~\ref{FigS6}(a) represents the dispersion map  along the $\mathrm{\overline{\Gamma}}-\mathrm{\overline{X}}$ direction obtained from the measurement using a photon energy of 90 eV. Bands crossing the Fermi level can be clearly observed, indicative of the absence of gap along this direction. In Supplementary~Figure~\ref{FigS6}(c), we show the dispersion map along the $\mathrm{\overline{M}}-\mathrm{\overline{X}}$ direction. Around the $ \mathrm{\overline{X}}$ point, bands are crossing the Fermi level, which again indicated the absence of any gap around the $\mathrm{\overline{X}}$ point as observed in the Fermi surface maps. The experimentally observed band dispersions along $\mathrm{\overline{\Gamma}}-\mathrm{\overline{X}}$ and  $\mathrm{\overline{M}}-\mathrm{\overline{X}}$ are well reproduced in the theoretical surface spectrum calculated without considering CDW ordering [Figures~\ref{FigS6}b,d].

\section*{\uppercase{Supplementary Note 7: Direction dependence of the CDW gap}}
In Supplementary~Figure~\ref{FigS7}, we present the dispersion maps, integrated EDCs and the Fermi fits of the leading edge for cut directions $\mathrm{\overline{\Gamma}}-\mathrm{\widetilde{A}}$ through $\mathrm{\overline{\Gamma}}-\mathrm{\widetilde{D}}$ as defined in Main text Figure 4. A clear signature of gap decreasing as the direction moves from $\mathrm{\overline{\Gamma}}-\mathrm{\widetilde{A}}$ towards $\mathrm{\overline{\Gamma}}-\mathrm{\widetilde{D}}$ can be observed. The results of the measurement taken at a temperature of $\mathrm{15~K}$ do not show significant differences (see Supplementary~Figure~\ref{FigS8}).\\~\\

\clearpage
\begin{figure*} [ht]
\centering
\includegraphics[width=1\textwidth]{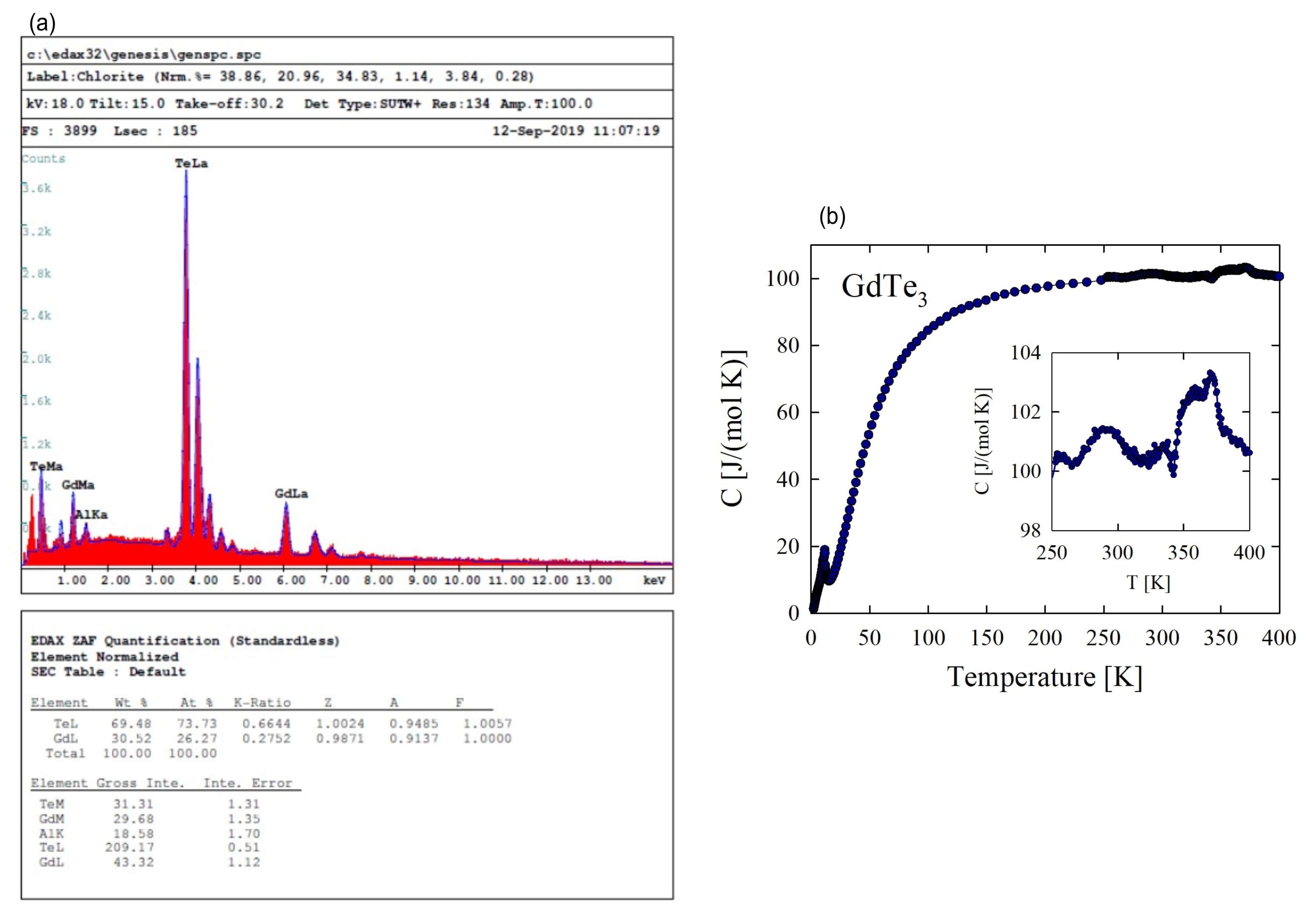}
\caption{\textbf{a} Microprobe analysis of the chemical composition of $\mathrm{Gd}\mathrm{Te}_3$ single crystal. \textbf{b} Temperature variation of the specific heat over a wide range of temperature. Inset: the data collected in the vicinity of the CDW transition.}
\label{FigS1}
\end{figure*}

\begin{figure*} [ht]
\centering
\includegraphics[width=0.95\textwidth]{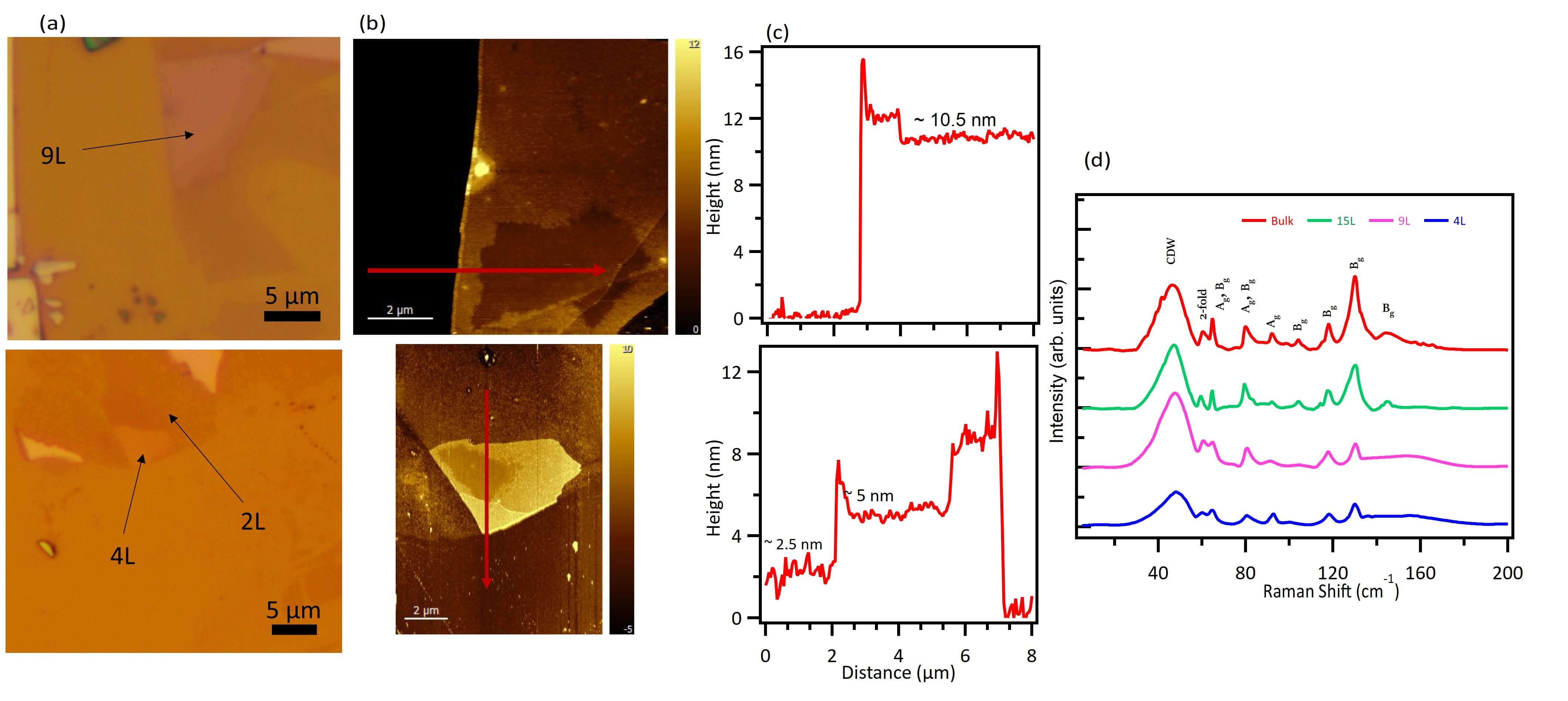}
\caption{Mechanical exfoliation of $\mathrm{Gd}\mathrm{Te}_3$ flakes and Raman spectroscopy results. \textbf{a} Optical image, \textbf{b} AFM image, and \textbf{c} height profiles of $\mathrm{Gd}\mathrm{Te}_3$ flakes exfoliated from bulk crystals. \textbf{d} Raman spectroscopy results on bulk and few layered samples indicated.}
\label{FigS2}
\end{figure*}

\begin{figure*} [ht]
\centering
\includegraphics[width=1\textwidth]{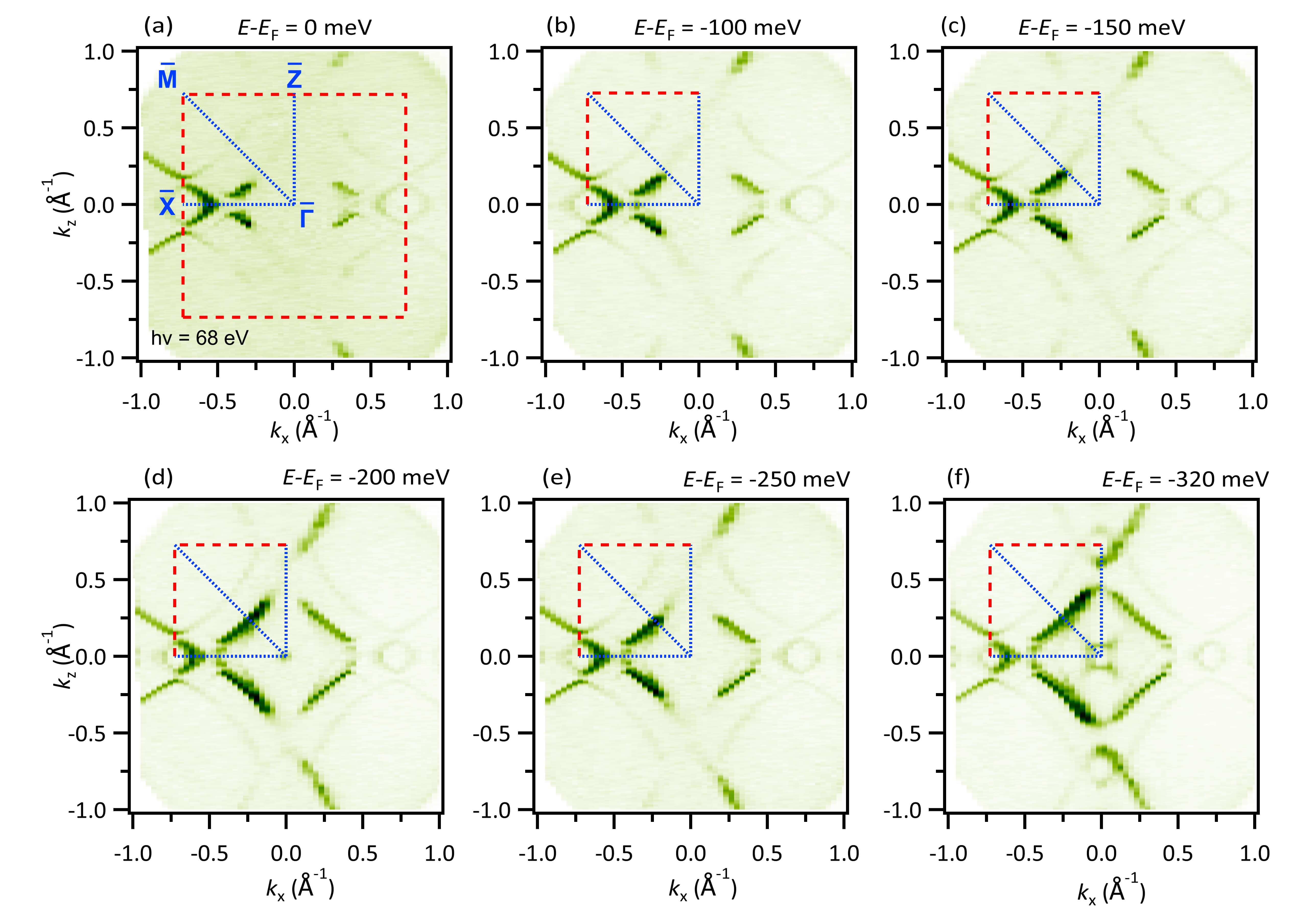}
\caption{Constant energy contours obtained for $\mathrm{68~eV}$ photon energy at binding energies of \textbf{a}~$\mathrm{0~meV}$ (FS), \textbf{b}~$\mathrm{100~meV}$ , \textbf{c}~$\mathrm{150~meV}$ , \textbf{d}~$\mathrm{200~meV}$ , \textbf{e}~$\mathrm{250~meV}$ , and \textbf{f}~$\mathrm{320~meV}$ (experimental Brillouin zone in green and high-symmetry lines/points in cyan color). Data were collected at the SSRL beamline 5-2 at a temperature of $\mathrm{8~K}$ .}
\label{FigS3}
\end{figure*}

\begin{figure*} [ht]
\centering
\includegraphics[width=1\textwidth]{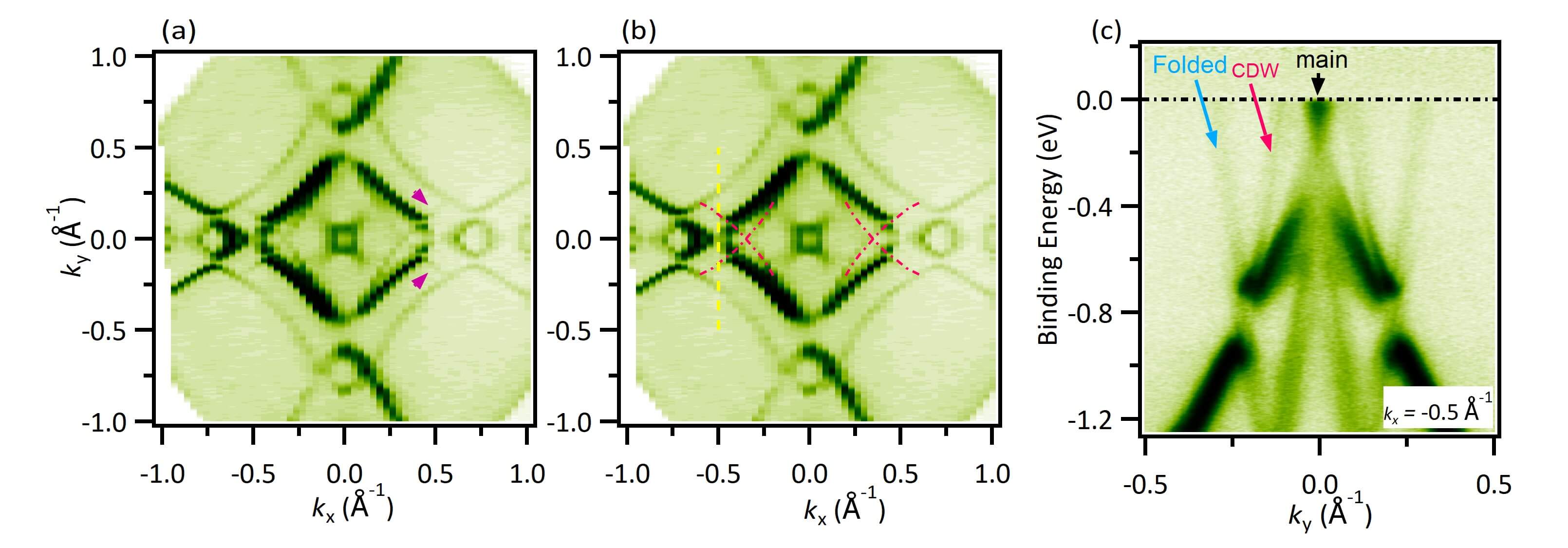}
\caption{Observation of CDW bands. \textbf{a} Energy contour integrated within a $\mathrm{50~meV}$ window centered at $\mathrm{300~meV}$ binding energy. \textbf{b} Same energy contour as in (a) with CDW induced features represented by magenta colored dashed curves. \textbf{c} Dispersion map taken for a $k_x=0~\AA^{-1}$ cut represented by the yellow dashed line in (b).}
\label{FigS4}
\end{figure*}

\begin{figure*} [ht]
\centering
\includegraphics[width=1\textwidth]{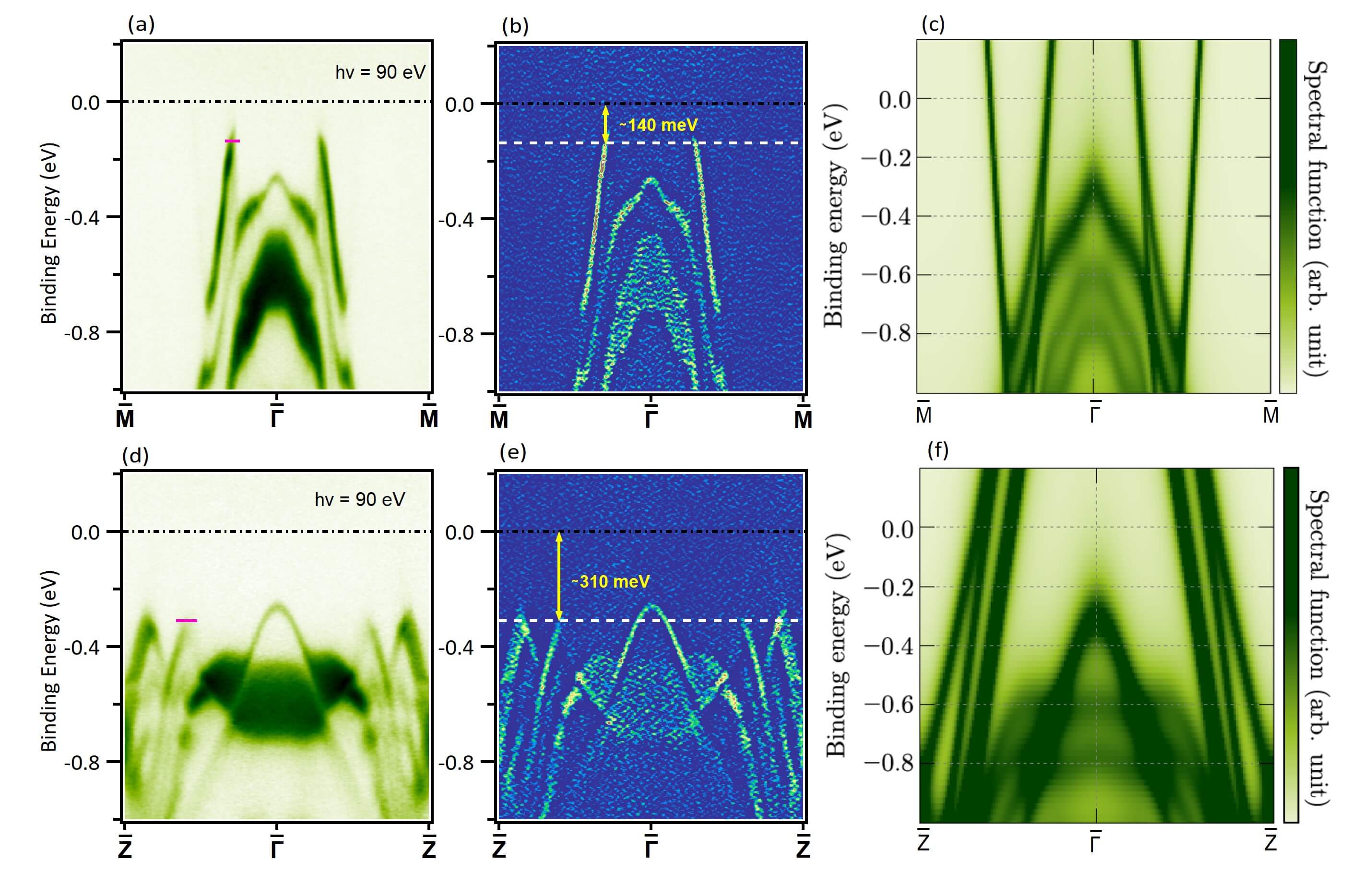}
\caption{Dispersion maps for 90 eV measurements. \textbf{a}~Dispersion map   and \textbf{b}~its second derivative along $\mathrm{\overline{M}}-\mathrm{\overline{\Gamma}}-\mathrm{\overline{M}}$. \textbf{c}~Calculated surface spectrum along $\mathrm{\overline{M}}-\mathrm{\overline{\Gamma}}-\mathrm{\overline{M}}$ without taking into account the CDW ordering. \textbf{d}~Dispersion map  and \textbf{e}~its second derivative  along $\mathrm{\overline{Z}}-\mathrm{\overline{\Gamma}}-\mathrm{\overline{Z}}$. \textbf{f}~Calculated surface spectrum along $\mathrm{\overline{Z}}-\mathrm{\overline{\Gamma}}-\mathrm{\overline{Z}}$ without considering CDW order.}
\label{FigS5}
\end{figure*}

\begin{figure*} [ht]
\centering
\includegraphics[width=1\textwidth]{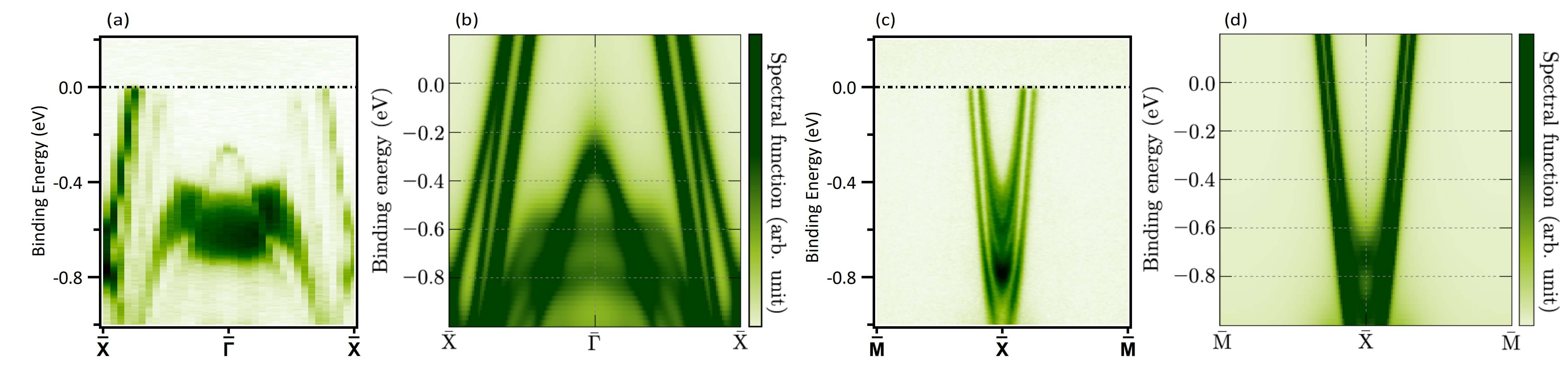}
\caption{Dispersion maps along $\mathrm{\overline{X}}-\mathrm{\overline{\Gamma}}-\mathrm{\overline{X}}$ and $\mathrm{\overline{M}}-\mathrm{\overline{X}} - \mathrm{\overline{M}}$.  \textbf{a}~Dispersion map along $\mathrm{\overline{X}}-\mathrm{\overline{\Gamma}}-\mathrm{\overline{X}}$. \textbf{b}~Calculated surface spectrum ($\mathrm{\overline{X}}-\mathrm{\overline{\Gamma}}-\mathrm{\overline{X}}$) projection on the (010) surface. \textbf{c}~Dispersion map along $\mathrm{\overline{M}}-\mathrm{\overline{X}} - \mathrm{\overline{M}}$. \textbf{d}~Calculated surface spectrum along $\mathrm{\overline{M}}-\mathrm{\overline{X}} - \mathrm{\overline{M}}$  projected on the (010) surface.}
\label{FigS6}
\end{figure*}

\begin{figure*} [ht]
\centering
\includegraphics[width=1\textwidth]{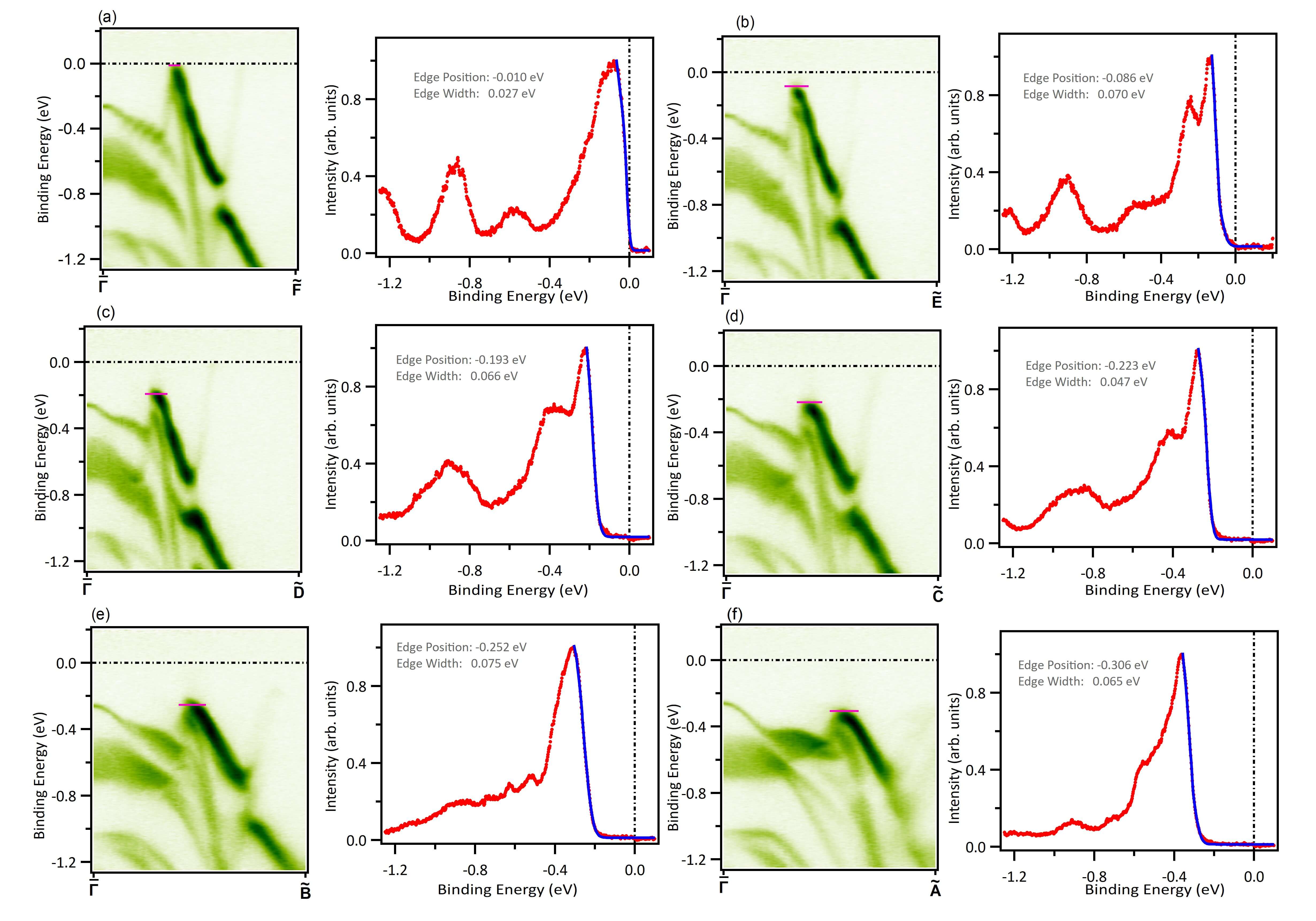}
\caption{Direction dependent gap. Dispersion maps (left panel) and the energy distribution curves integrated within the momentum window indicated by the magenta lines in the dispersion maps (right panel) along \textbf{a}~$\overline{\mathrm{\Gamma}}-\widetilde{\mathrm{F}}$, \textbf{b}~$\overline{\mathrm{\Gamma}}-\widetilde{\mathrm{E}}$, \textbf{c}~$\overline{\mathrm{\Gamma}}-\widetilde{\mathrm{D}}$, and \textbf{d}~$\overline{\mathrm{\Gamma}}-\widetilde{\mathrm{C}}$, \textbf{e}~$\overline{\mathrm{\Gamma}}-\widetilde{\mathrm{B}}$, \textbf{f}~$\overline{\mathrm{\Gamma}}-\widetilde{\mathrm{A}}$. These directions represent the cuts at an angle of respectively: $\mathrm{62\degree}$, $\mathrm{55\degree}$, $\mathrm{40\degree}$, $\mathrm{30\degree}$, $\mathrm{20\degree}$, and $\mathrm{10\degree}$, from the $\mathrm{\overline{\Gamma}}-\mathrm{\overline{Z}}$ direction. The leading edges have been fitted by Fermi edge fitting.}
\label{FigS7}
\end{figure*}

\begin{figure*} [ht]
\centering
\includegraphics[width=0.75\textwidth]{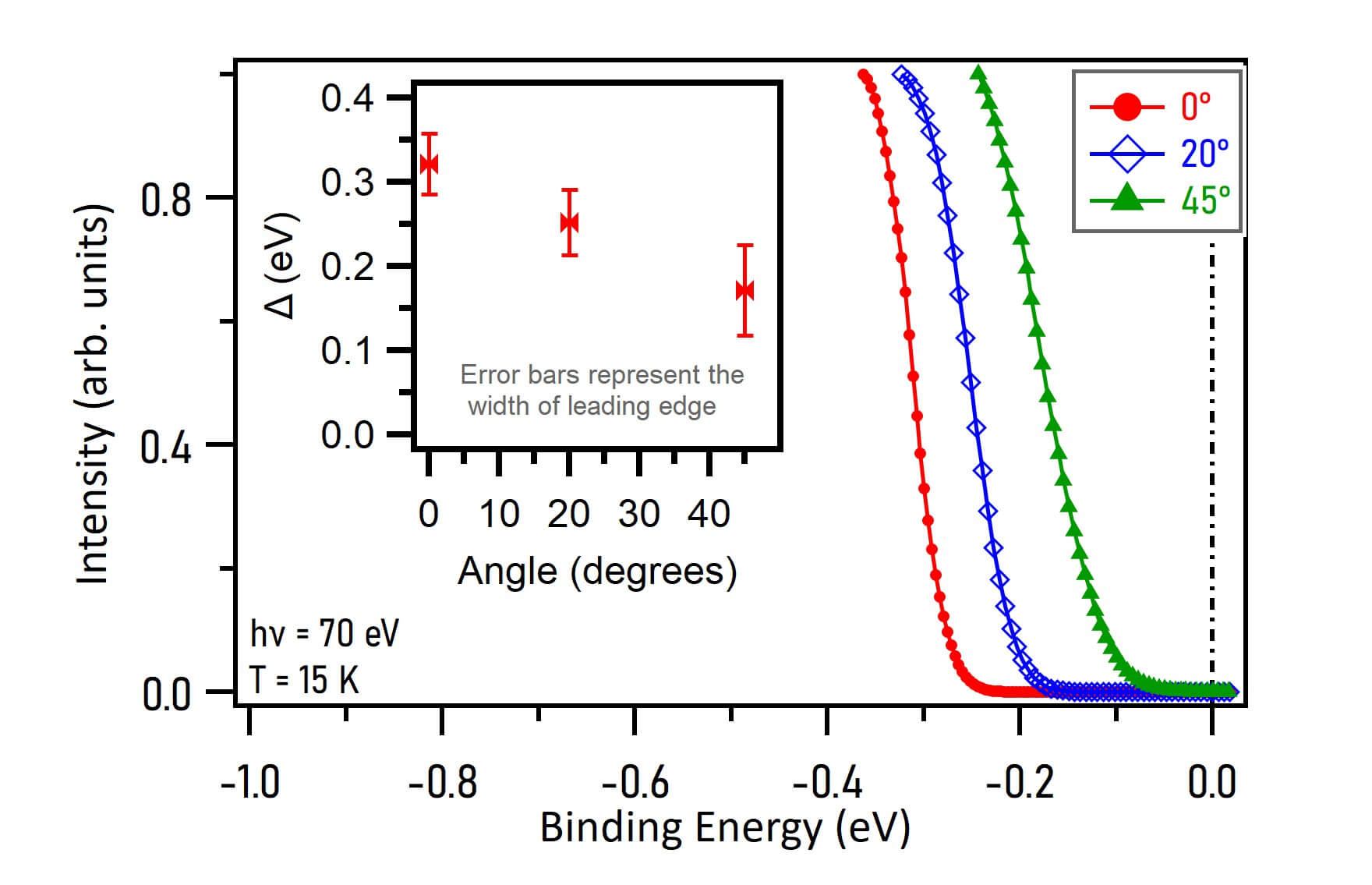}
\caption{Fitted leading edges (normalized) for the angles noted in the plot. Inset: Gap size below the Fermi level, where the error bars represent the width of the leading edge. Data were taken at the ALS beamline 10.0.1.2 at a temperature of $\mathrm{15~K}$. }
\label{FigS8}
\end{figure*}

\end{document}